\documentclass[conference]{IEEEtran}
\ifCLASSINFOpdf
\usepackage[pdftex]{graphicx}
\else
\fi
\usepackage{algorithm} 
\usepackage{algpseudocode}
\usepackage{url}

\begin{document}
%
\title{A Heuristic Method of Generating Diameter 3 Graphs for Order/Degree Problem}

\author{\IEEEauthorblockN{Teruaki Kitasuka}
\IEEEauthorblockA{Graduate School of Science and Technology\\
Kumamoto University\\
Kumamoto 860-8555, Japan\\
Email: kitasuka@cs.kumamoto-u.ac.jp}
\and
\IEEEauthorblockN{Masahiro Iida}
\IEEEauthorblockA{Graduate School of Science and Technology\\
Kumamoto University\\
Kumamoto 860-8555, Japan\\
Email: iida@cs.kumamoto-u.ac.jp}
}


%


\IEEEoverridecommandlockouts
\IEEEpubid{}

\maketitle

\begin{abstract}
We propose a heuristic method that generates a graph for order/degree
problem.  Target graphs of our heuristics have large order ($>$ 4000)
and diameter 3.  We describe the observation of smaller graphs and
basic structure of our heuristics.  We also explain an evaluation
function of each edge for efficient 2-opt local search.  Using them,
we found the best solutions for several graphs.
\end{abstract}

\def\IEEEkeywordsname{Keywords}

%
\IEEEpeerreviewmaketitle

\section{Introduction}
One of the famous problems in the field of combinatorics is the
degree/diameter problem~\cite{Erdos1980,Loz2008,Miller2013,Exoo2013}.
The degree/diameter problem\footnote{The Degree/Diameter Problem,
  CombinatoricsWiki,
  \url{http://combinatoricswiki.org/wiki/The_Degree/Diameter_Problem#Undirected_graphs}}
is the problem of finding the largest possible number $n(d, k)$ of
nodes in a graph of maximum degree $d$ and diameter $k$.  The maximum
degree of a graph is the maximum degree of its nodes.  The degree of a
node is the number of edges incident to the node.  The diameter $k$ of
a graph is the maximum distance between two nodes of the graph.

On the other hand, the problem of the graph golf 2015
competition~\cite{GG2015} is the order/degree problem.  The
order/degree problem is the problem of finding a graph that has
smallest diameter $k$ and average shortest path length (ASPL, $l$) for
a given order and degree.  Compared to the degree/diameter
problem, order is given and diameter is not given in the order/degree
problem.

As the organizer of the competition mentioned, the order/degree
problem has important role to design networks for high perfomance
computing.  Because, the number of nodes of the network is determined
based on design constraints such as cost, space, budget, and
applications.  Solutions of the degree/diameter problem can be used to
limited networks of particular number of nodes.  Currently, there is
no trivial way to increase or decrease the number of nodes from the
optimal graph of the degree/diameter problem, while keeping its
diameter close to the optimal graph.  For example, Besta and
Hoefler~\cite{Besta2014} have presented diameter-2 and -3 networks
with particular number of routers, and each endpoint is connected to a
router.  The number of endpoints can be changed in a range.  Matsutani
et al.~\cite{Matsutani2014} have reduced the communication latency of
3D NoCs, by adding randomized shortcut links.

We try to solve some order/degree problems.  There are two
contributions in this paper.  1) Showing heuristic algorithm to create
a graph for given order and degree (Sec.~\ref{sec:algorithm}).  Using
this algorithm, we have created two best-known graphs; one has order
$n = 4096$ and degree $d = 60$, the other has $n = 4096$ and $d = 64$.
After 2015 competition, we also have created other graph of order $n =
10000$ and degree $d = 60$.
2) Developing evaluation function of edges for 2-opt local search
(Sec.~\ref{sec:search}).  Local search starts with a graph that has
the given number of nodes and satisfies degree constraints.  Swapping
two edges is accepted, if swapped graph $G'$ is better than the
previous graph $G$ in terms of diameter and/or ASPL.  For example, if
two edges $a$-$b$ and $c$-$d$ are selected for swapping from graph
$G$, we try to swap two edges such that two edges $a$-$b$ and $c$-$d$
are removed from and two edges $a$-$c$ and $b$-$d$ are added to the
graph $G$.  If diameter and/or ASPL of swapped graph $G'$ is smaller
than $G$, this swap is accepted.  We call the evaluation function
``edge importance''.  Lower-importance edge pair is selected as the
candidate of swapping earlier than other pairs.  For the existence of
local minimum graph, we need to temporarily accept worse graphs in
searching graph of order $n = 256$ and degree $d = 16$.


\section{Observation of Small Order Graphs}
\label{sec:small}

\begin{figure}[tb]
  \begin{center}
    \includegraphics[scale=.5]{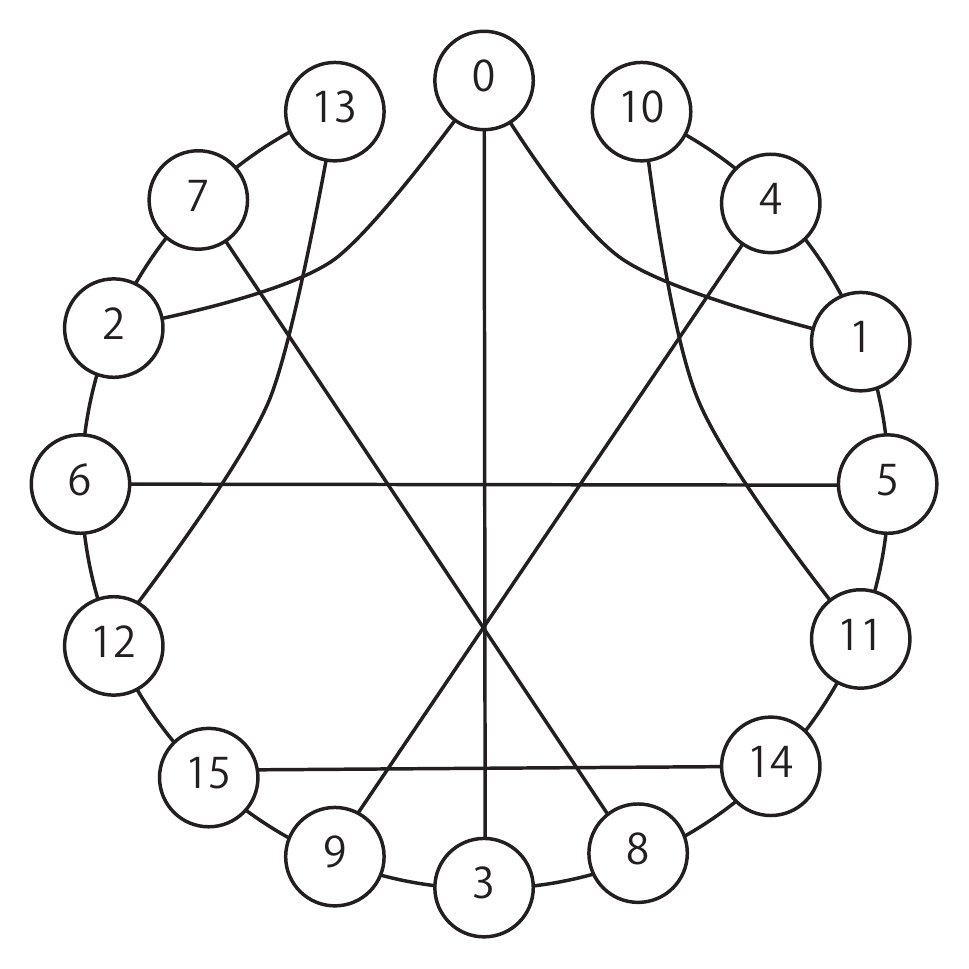}

    \footnotesize
    (a) ring layout
    
    \includegraphics[scale=.5]{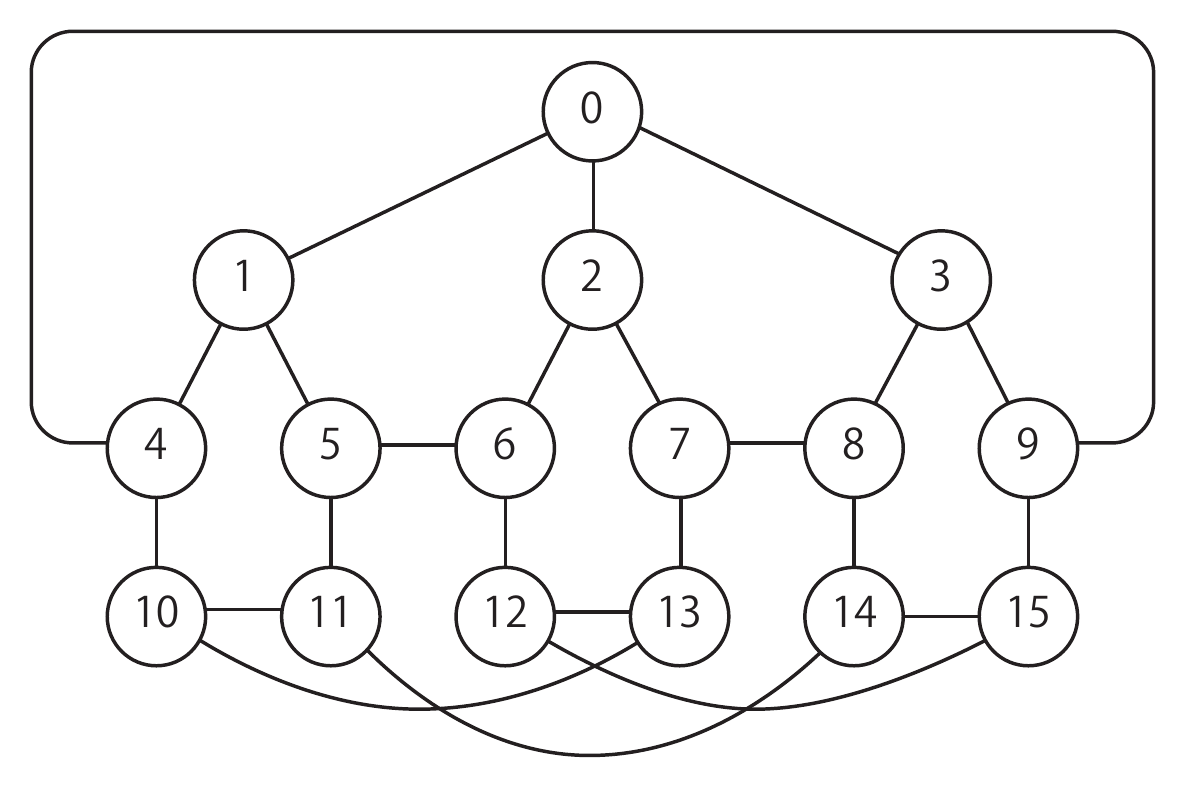}

    (b) pentagon (5-node cycle) layout
  \end{center}
  \caption{The best known graph with order $n = 16$ and degree $d = 4$}
  \label{fig:n16d3zdd5}
\end{figure}

\begin{figure}[tb]
  \begin{center}
    \includegraphics[scale=.5]{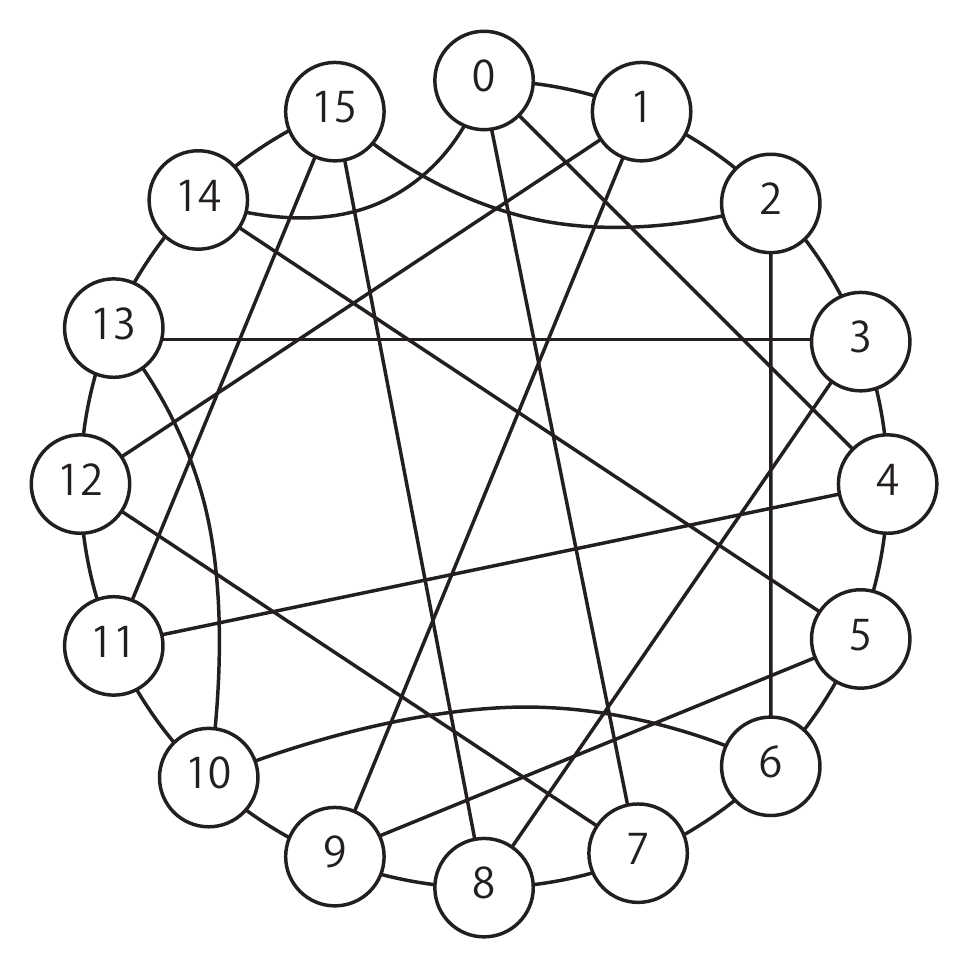}
    \footnotesize

    (a) ring layout
    
    \includegraphics[scale=.5]{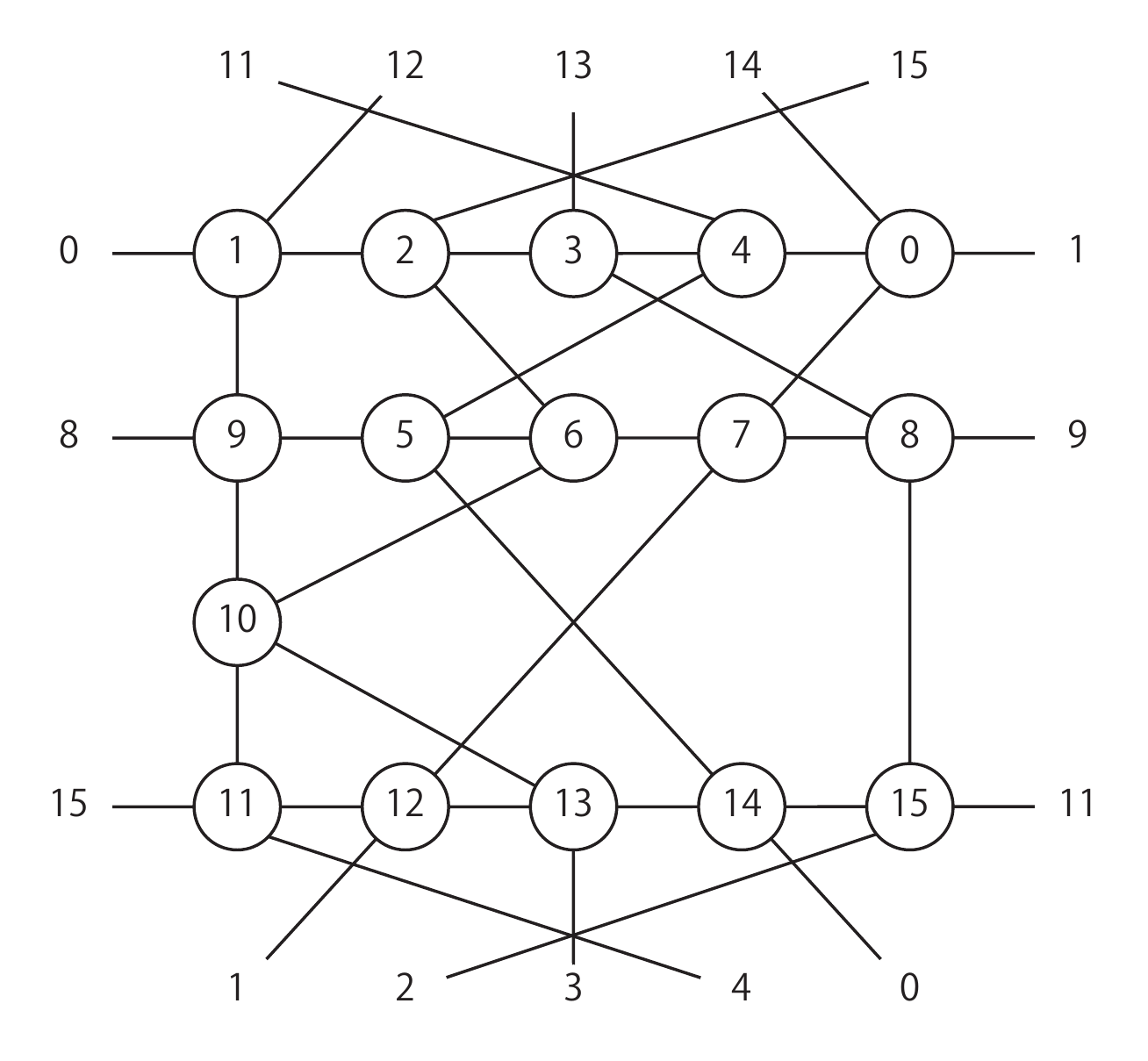}

    (b) pentagon (5-node cycle) layout

    \includegraphics[scale=.5]{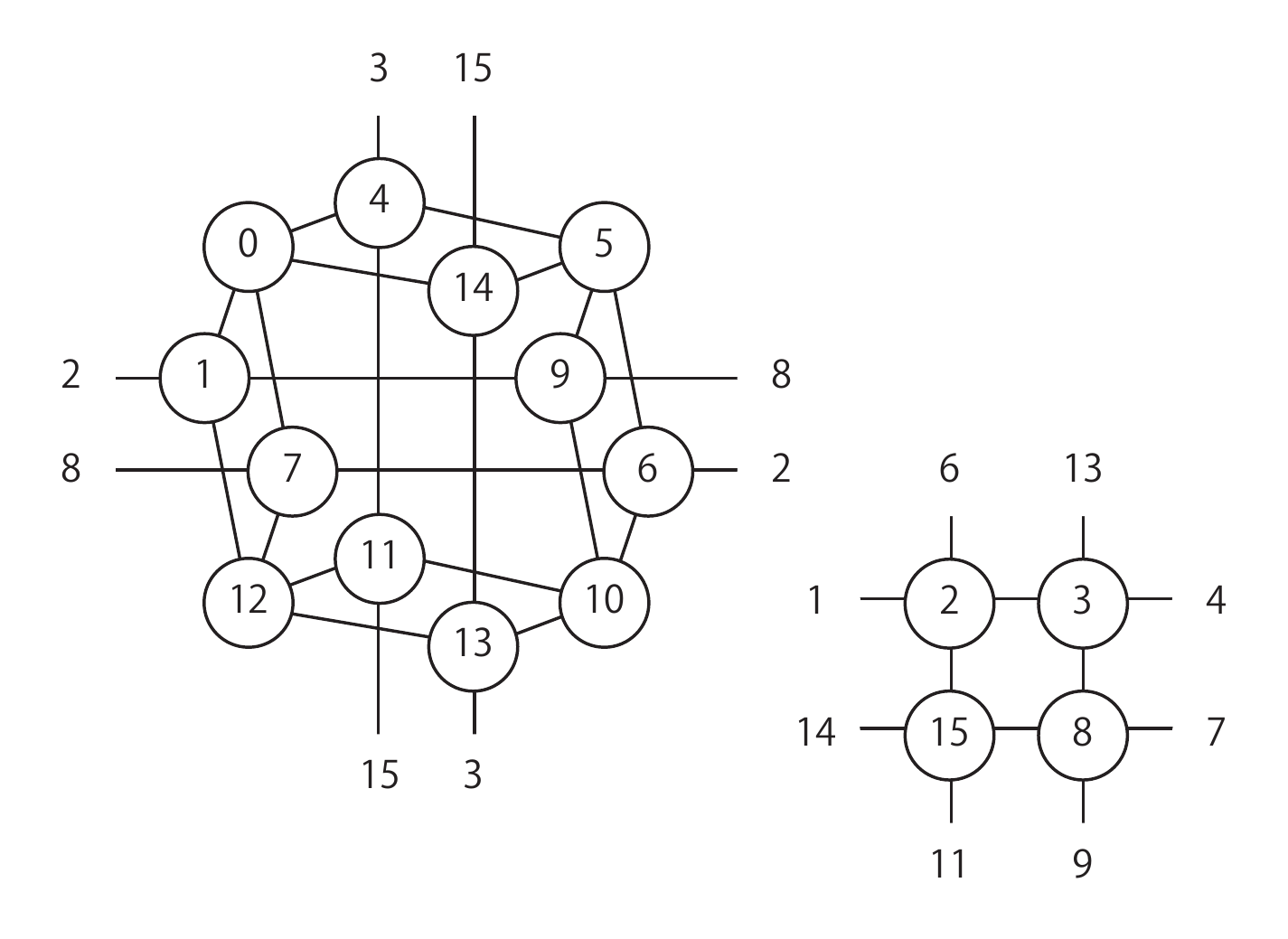}

    (c) square (4-node cycle) layout

  \end{center}
  \caption{The best known graph with order $n = 10$ and degree $d = 3$}
  \label{fig:n16d4iida13}
\end{figure}

The observation of small order graphs brings us the idea of the
heuristic algorithm shown in Sec.~\ref{sec:algorithm}.
At the beginning of the 2015 competition, we drew graphs with small
order and degree.  The first graph is order $n = 16$ and degree $d =
3$ as shown in Fig.~\ref{fig:n16d3zdd5}.  The second one is order $n =
16$ and degree $d = 4$ as shown in Fig.~\ref{fig:n16d4iida13}.

The diameters of these two graphs are three ($k = 3$).  Through
drawing these two graphs, we found that these graphs contain many
pentagons (5-node cycles), no or small number of squares (4-node
cycles), and no triangle (3-node cycle).  In Fig.~\ref{fig:n16d3zdd5},
there is no triangle and no square.  No triangle and four squares
exist in Fig.~\ref{fig:n16d4iida13}.  We think triangles and squares
cause diameter ASPL (average shortest path length) to be larger for
the case of $k = 3$.  Through this observation, we define increasing
the number of pentagons as our policy in Section \ref{sec:algorithm}.
In the degree/diameter problem, pentagons are appeared in the graphs
of diameter $k = 2$, e.g., Petersen graph (shown in
Fig.~\ref{fig:petersenGraph}) and Hoffman-Singleton graph ($n = 50$
and $d = 7$).

\section{Heuristic Algorithm}
\label{sec:algorithm}

\subsection{Policy and Outline of Heuristic Algorithm}

Based on the observation of small order graphs described in
Sec.~\ref{sec:small}, we determine the outline of our heuristic
algorithm as following two steps.  1) If target diameter is $k$, we
connect small order graphs such that their diameter is $k - 1$.  For
example, if the target diameter $k = 3$ and order $n = 10000$, the
$1000$ Petersen graphs (Fig.~\ref{fig:petersenGraph}, diameter $k =
2$) are connected.  2) We try to increase the number of $(2k-1)$-node
cycles, when edges are added.  For graphs of $k=3$, we try to increase
the number of pentagons (5-node cycles).

Outline of our heuristic algorithm is shown in Algorithm \ref{alg:gg}.
In the remaining of the paper, we discuss only $k=3$ graphs.  Our
algorithm generates a graph which diameter is almost 3, for given
order and degree.

\begin{algorithm}
  \caption{Outline of graph generation algorithm}\label{alg:gg}
  \begin{algorithmic}[1]
    \Procedure{GraphGenerater}{order $n$, degree $d$}
      \State Create a base graph $G_0$, such that order of $G_0$ is $n$
      \State Create a graph $G$, by greedily adding edges one by one to $G_0$
      \State Return the graph $G$
    \EndProcedure
  \end{algorithmic}
\end{algorithm}

\subsection{Create a Base Graph $G_0$}

A base graph $G_0 = (V, E_0)$ has $n$ nodes, i.e. $|V| = n$.  The
graph $G_0$ is a connected graph, but its degree is five.  Most nodes
have five edges.  Other nodes, i.e., some border and anomalous nodes
have four edges.

\begin{figure}[tb]
  \begin{center}
    \includegraphics[scale=.5]{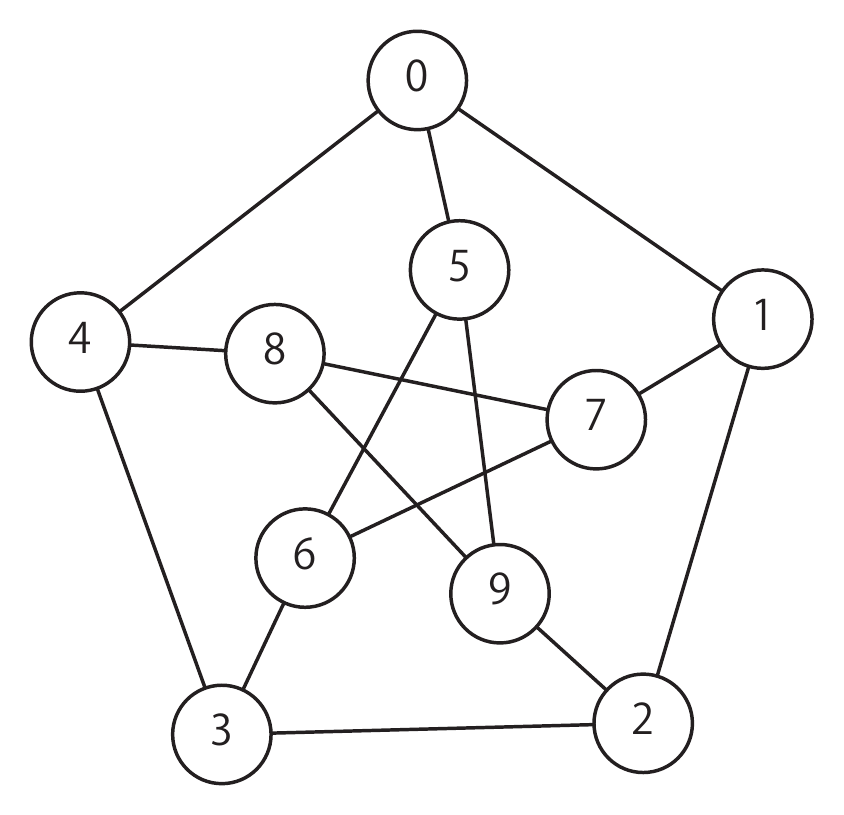}
  \end{center}
  \caption{The Petersen graph $(n=10, d=3)$}
  \label{fig:petersenGraph}
\end{figure}

Graph $G_0$ contains multiple Petersen graphs.  The Petersen graph
$G_P$, which is shown in Fig.~\ref{fig:petersenGraph}, is one of
well-known Moore graphs~\cite{Miller2013}, and has ten nodes and
degree $d = 3$.  The diameter of Petersen graph is two.  When the
nodes are numbered in Fig.~\ref{fig:petersenGraph}, fifteen edges of
the Petersen graph are described as follows, for $i \in \{0, 1, 2, 3,
4\}$ and $j \in \{5, 6, 7, 8, 9\}$.

\def\mod{{\rm \;mod\; }}
\begin{eqnarray*}
  && (i, i + 1 \mod 5) \\
  && (i, (2 i \mod 5) + 5) \\
  && (j, (j + 1 \mod 5) + 5)
\end{eqnarray*}

\begin{figure}[tb]
  \begin{center}
    \includegraphics[scale=.5]{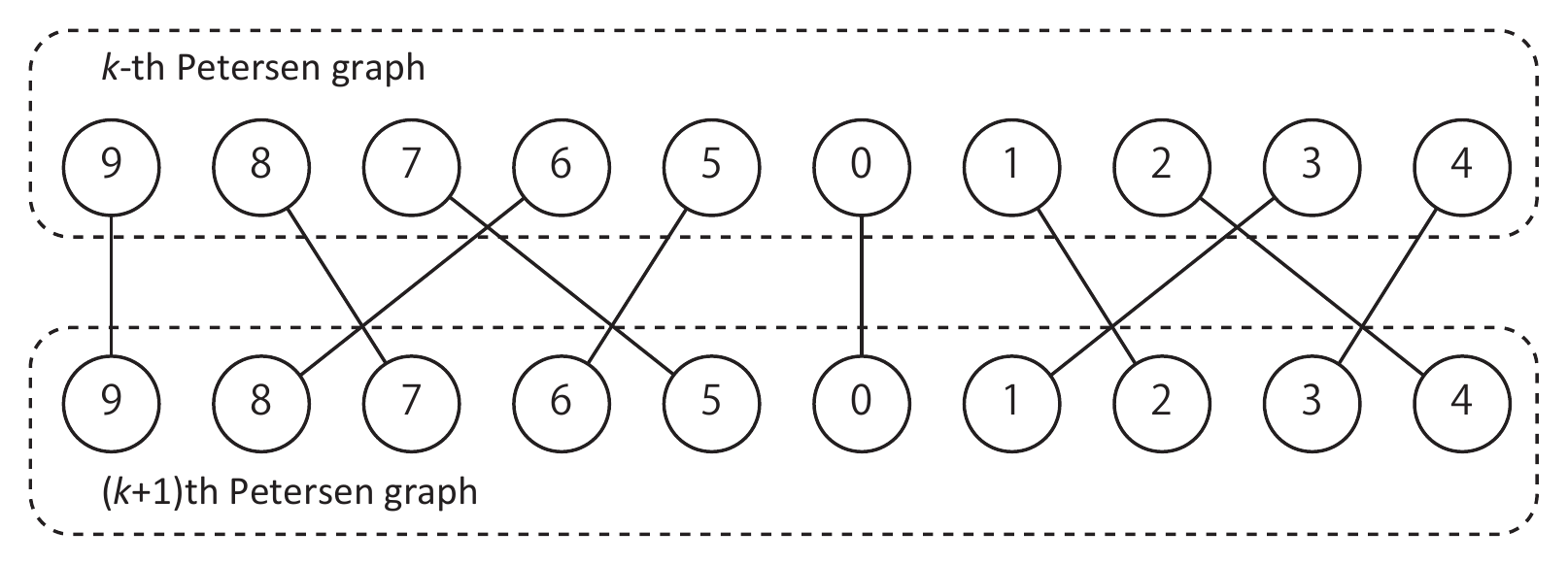}
  \end{center}
  \caption{Connecting two adjacent Petersen graphs}
  \label{fig:connectPetersenGraph}
\end{figure}

If a given order $n$ is multiple of ten, we generate $(n / 10)$
Petersen graphs, $G_{Pk} \;(k = 1, 2, 3, \ldots, n / 10)$.  Adjacent
Petersen graphs, $G_{Pk}$ and $G_{P(k+1)}$, are connected as shown in
Fig.~\ref{fig:connectPetersenGraph} by {\sc Connect} procedure in
Algorithm~\ref{alg:baseGraph}.  Fig.~\ref{fig:connectPetersenGraph}
shows only edges crossing two Petersen graphs.  In the case of $n =
10000$, we generate 1000 Petersen graphs, and $k$-th graph are
connected with $(k-1)$-th and $(k+1)$-th graph for $1 < k < 1000$.

\begin{algorithm}
  \caption{Create a base graph $G_0$}
  \label{alg:baseGraph}
  \begin{algorithmic}[1]
    \Procedure{CreateBaseGraph}{$n$}
      \For{\textbf{each} $k \in \{1, \ldots, (n / 10)\}$}
        \State Create $k$-th petersen graphs
      \EndFor
      \For{\textbf{each} $k \in \{1, \ldots, (n / 10) - 1\}$}
        \State {\sc Connect}$(k)$
      \EndFor
    \EndProcedure
    \Procedure{Connect($k$)}{}
    \State Connect $k$-th and $(k+1)$-th Petersen graphs as shown in Fig.~\ref{fig:connectPetersenGraph}
    \EndProcedure
  \end{algorithmic}
\end{algorithm}

\begin{figure}[tb]
  \begin{center}
    \includegraphics[scale=.5]{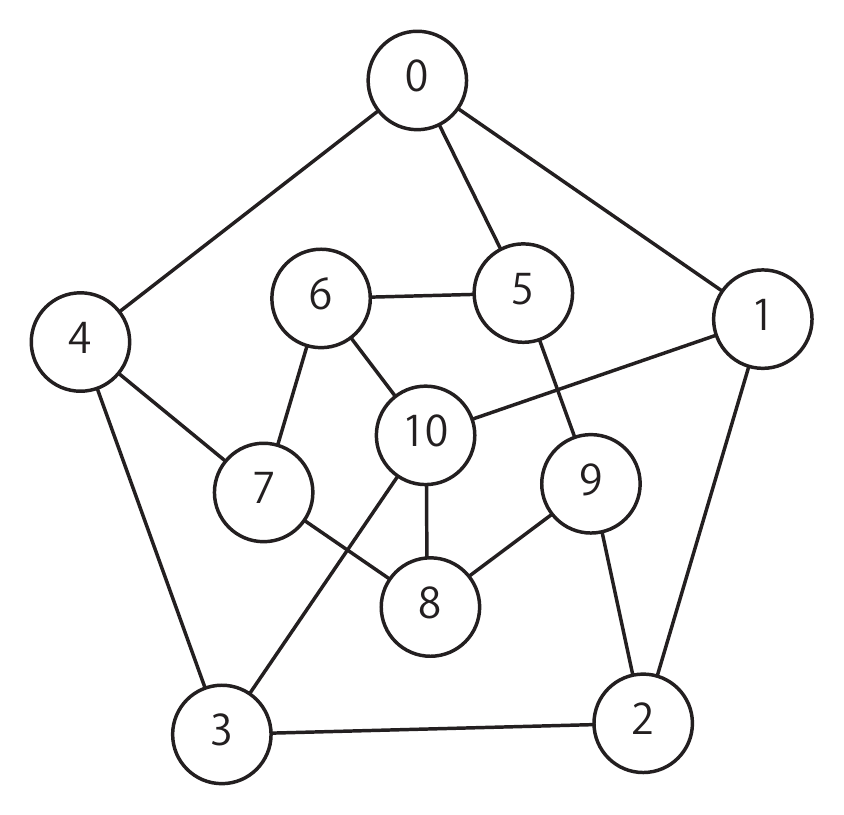}
  \end{center}
  \caption{The 11-node graph $(n=11, d=4)$ for base graphs}
  \label{fig:11nodesGraph}
\end{figure}

When there is a remainder $r > 0$ divided by ten, i.e., $n = 10 \cdot
\lfloor n / 10 \rfloor + r$, we replace $r$ Petersen graphs with $r$
11-node graphs.  The 11-node graph is the subgraph of
Fig.~\ref{fig:n16d4iida13}.  We heuristically select eleven nodes, $5,
6, 7, \ldots, 15$ from the graph of Fig.~\ref{fig:n16d4iida13}.  When
we connect a 11-node graph with adjacent Petersen graphs, we ignore
node 10 and other ten nodes are connected similar to
Fig.~\ref{fig:connectPetersenGraph}.  The eleven nodes graph is shown
in Fig.~\ref{fig:11nodesGraph}, nodes are renumbered, except for node
10.  Nodes $5, 6, 7, 8, 9, 11, 12, 13, 14, 15$ in
Fig.~\ref{fig:n16d4iida13} are renumbered to $2, 3, 4, 0, 1, 6, 7, 8,
9, 5$ in Fig.~\ref{fig:11nodesGraph}, respectively.

The base graph $G_0$ is generated by {\sc CreateBaseGraph} procedure
in Algorithm~\ref{alg:baseGraph}.  The base graph $G_0$ has $n$ nodes.
Each node of $G_0$ has five edges, except for nodes in the first and
the last Petersen graphs and node 10 of 11-node graphs.  These
exceptional nodes have just four edges.
  
\subsection{Greedily Add Edges One by One to $G_0$}

In this step, we greedily add edges one by one to the base graph
$G_0$.  Our policies to add edges are the followings.
\begin{enumerate}
\item Increase the number of pentagons in the graph, to create a graph
  such that its diameter becomes three and its ASPL is close to two.
\item Add an edge, which has the smallest degree node on one side.
\item No track back, i.e., never remove edges from the graph.
\end{enumerate}

Under the policy 1), our heuristic searches two nodes such that
distance of them is four, and adds an edge between these two nodes.
By adding the edge, the number of pentagons is increased.  Even if the
small-degree graph of $n = 16$ and $d = 4$ in
Fig.~\ref{fig:n16d4iida13}, there are many pentagons those include a
particular edge.  For example, an edge 1-2 is contained in eight
pentagons; 1-2-3-4-0, 1-2-6-7-0, 1-2-15-14-0, 1-2-3-8-9, 1-2-3-13-12,
1-2-6-5-9, 1-2-6-7-12, and 1-2-6-10-9.

The policy 2) is employed to uniformly increase the degree of nodes
and save computation time.  Our heuristic maintains the nodes that
have the smallest degree, and selects a node from them as one side of
a new edge.  A node of the other side is selected based on the policy
1).  Although we can select the new edge from all possible pair of
nodes, to save computation time, our heuristic limits search space by
fixing one side of new edge.

The policy 3) also saves computation time.  As another reason, we do
not find any effective evaluation function to track back.

Here, we explain this step in Algorithm~\ref{alg:addEdges}.  In each
loop iteration of lines~\ref{l:while1} to \ref{l:while1end}, two nodes
$i$ and $j$ are selected and add an edge $i$-$j$ to the graph.  Node
$i$ is chosen from the nodes of the smallest degree in $G$ at
line~\ref{l:selecti}, based on policy 2).  $d(i, j)$ denotes the
distance between two nodes $i$ and $j$.  In the loop of
lines~\ref{l:for1} to \ref{l:for1end}, candidates of node $j$ are
evaluated, based on policy 1).  After evaluation, node $j$ that
satisfies two following conditions (\ref{eqn:p1}) and (\ref{eqn:p2})
is selected in line~\ref{l:selectj}, and an edges $i$-$j$ is added to
graph $G$ in the next line.  $J'$ is the subset of $J$ such that each
node in $J'$ satisfies condition (\ref{eqn:p1}).  We have no
particular tie-breaking rule.
\begin{eqnarray}
  p_1(j) &=& \min_{j'\in J} p_1(j') \label{eqn:p1} \\
  p_2(j) &=& \max_{j'\in J'} p_2(j') \label{eqn:p2}
\end{eqnarray}

The {\sc CountPaths} function is used for the evaluation of $j \in J$.
{\sc CountPaths}$(i, j)$ roughly counts the number of paths between
two nodes $i$ and $j$, those distance are three.  $D_m(i)$ is the set
of nodes distant $m$ from node $i$, i.e., for every node $k \in
D_m(i)$, $k$ satisfies $d(i, k) = m$.  For example, every node $k \in
D_1(i) \cap D_2(j)$ satisfy $d(i, k) = 1$ and $d(k, j) = 2$.  $p$ in
line~\ref{l:countpathp} is close to the twice of the number of 4-node
paths.

\begin{algorithm}
  \caption{Greedily add edges, one by one to $G_0$}
  \label{alg:addEdges}
  \begin{algorithmic}[1]
    \Procedure{AddEdges}{$n, d, G_0$}
      \State $G \gets G_0$
      \While{edge can be added}\label{l:while1}
        \State\label{l:selecti} Select a node $i$ from the smallest degree nodes
        \State Compute node set $J$ such that $d(i, j) > 2$
        \For{\textbf{each} node $j \in J$}\label{l:for1}
          \State $p_1(j) = \mbox{\sc CountPaths}(i, j)$
          \State $p_2(j) = 0$
          \For{\textbf{each} $k \in j$'s neighbors}
            \If{$\mbox{\sc CountPaths}(i, k) > p_2(j)$}
              \State $p_2(j) = \mbox{\sc CountPaths}(i, k)$
            \EndIf
          \EndFor
        \EndFor\label{l:for1end}
        \State Select $j \in J$ that satisfies conditions (\ref{eqn:p1}) and (\ref{eqn:p2}) \label{l:selectj}
        \State Add an edge $i$-$j$ to graph $G$
      \EndWhile\label{l:while1end}
      \State If degree of several nodes are less than $d$, add several
      edges between them
    \EndProcedure

    \Function{CountPaths}{$i, j$}
      \State $p = | D_1(i) \cap D_2(j) | + | D_2(i) \cap D_1(j) |$
      \label{l:countpathp}
      \Comment Roughly count the number of paths with distance 3
      between node $i$ and $j$, for graph $G$
    \EndFunction
  \end{algorithmic}
\end{algorithm}

\subsection{Generated Graphs}

Diameter $k$ and ASPL $l$ of generated graphs are shown in
Table~\ref{tbl:heuristicResults}.  Fortunately, two graphs of $n =
4096$ are the new records in the competition.  The graph of $n = 256$
has the same diameter, but longer ASPL than the best record $l =
2.09262$ and two competitors' records.
For two graphs of $n = 10000$, our first implementation is too slow
and can not finish before the deadline of the 2015 competition.  After
the competition, we reimplement the program and get the results as
shown in Table~\ref{tbl:heuristicResults}.

\begin{table}
  \caption{Graphs generated by the heuristic algorithm}
  \label{tbl:heuristicResults}
  \centering
  \begin{tabular}{rr|rr|l}
    order $n$ & degree $d$ & diameter $k$ & ASPL $l$ & note \\
    \hline
    256 & 16 & 3 & 2.12757 & not submitted \\ 
    4096 & 60 & 3 & 2.295275 & *1\\ 
    4096 & 64 & 3 & 2.242228 & *1\\ 
    10000 & 60 & 3 & 2.648980 & *2 \\ 
    10000 & 64 & 3 & 2.611310 & *3 \\ 
    \\
    \multicolumn{5}{r}{\begin{minipage}{.7\linewidth}

        *1: None submits smaller graph in the competition.
        
        *2: A smaller-ASPL graph is created after competition.  The
        winner's graph of the competition has $l = 2.650157$.

        *3: The graph is created after competition.  However, the
        winner's graph of the competition ($l = 2.609927$) has smaller
        ASPL than this.
    \end{minipage}}
  \end{tabular}
\end{table}

\section{A Technique of 2-Opt Local Search}
\label{sec:search}

\subsection{Edge Importance Function}

After a graph is created by heuristic algorithm described in
Sec.~\ref{sec:algorithm} for a given order and degree, we start 2-opt
local search.  2-opt is the basic and widely used local search
heuristic~\cite{Englert2007}.  It is used for traveling salesperson
problem (TSP) and others.

\begin{figure}[tb]
  \begin{center}
    \includegraphics[scale=.5]{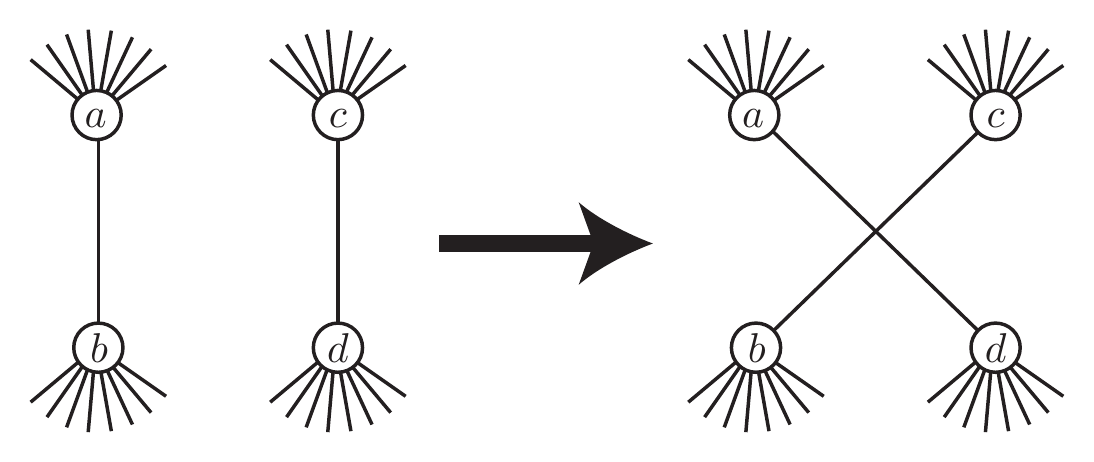}
  \end{center}
  \caption{Swap two edges $a$-$b$ and $c$-$d$ into $a$-$d$ and
    $b$-$c$.  (We assume degree $d = 9$ and other nodes are not drawn.
    There is another swap into $a$-$c$ and $b$-$d$ for these edges)}
  \label{fig:swapExmaple}
\end{figure}

In this section, we explain edge importance which is used to
prioritize edge combinations for 2-opt local search.  2-opt algorithm
slightly modifies a given graph recursively.  The modification of
2-opt is swapping two edges.  An example of swapping two edges $a$-$b$
and $c$-$d$ into $a$-$d$ and $b$-$c$ is shown in
Fig.~\ref{fig:swapExmaple}.  Diameter and ASPL of pre-swap graph $G =
(V, E)$ and post-swap graph $G' = (V, E')$ are compared with each
other.  If diameter and/or ASPL of $G'$ is smaller than $G$, this swap
is accepted.

2-opt local search is time-consuming task.  There are many ways to
reduce computation time.  Even if we search a graph of order $n = 256$
and degree $d = 16$, there are 2048 edges in the graph.  The number of
edge pairs is about $2\times10^6$.  Generally, the number of edge
pairs is about $\frac{nd}{2}\cdot\frac{(n - 2)d}{2}\cdot\frac{1}{2} =
O(n^2d^2)$.  For each swapped graph, we need to calculate diameter and
ASPL.  We adopt two techniques to save computation time.  One is edge
importance, and the other is fast ASPL calculation for 2-opt.

Our edge importance (or edge impact) is a value given to each edge of
a graph.  As an intuitive explanation, less important edges probably
be removed from the graph with little increase of ASPL than other
edges.  Then, we give higher priority to less important edges, when we
select an edge pair for swap.

\begin{figure}[tb]
  \begin{center}
    \includegraphics[scale=.45]{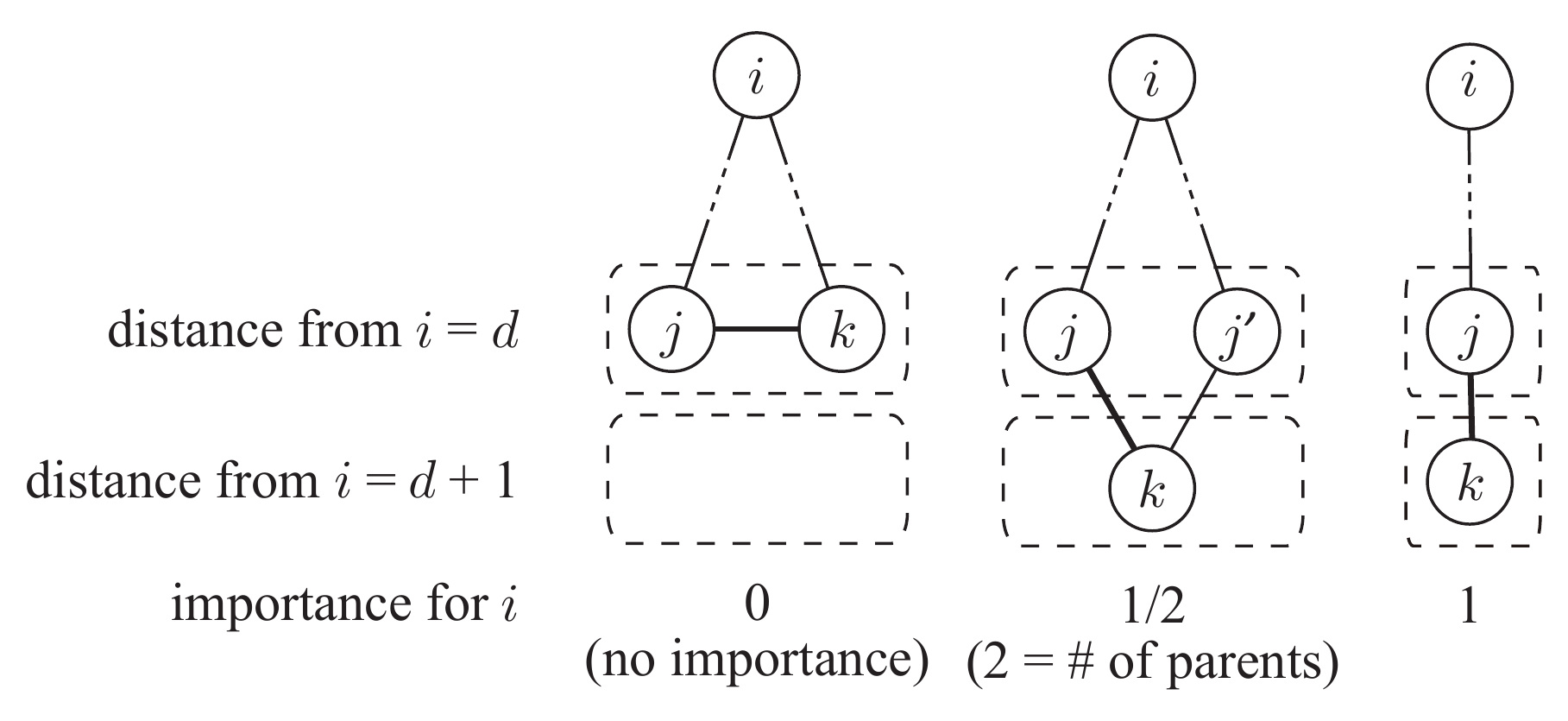}
  \end{center}
  \caption{Edge importance of an edge $j$-$k$ for node $i$}
  \label{fig:edgeImportance}
\end{figure}

The edge importance of an edge $e = j$-$k$ is defined by the following
function.
\[
f(e) = f(j\mbox{-}k) = \sum_{i \in V} f_1(i, j\mbox{-}k),
\]
where $f_1(i, j\mbox{-}k)$ is the importance of edge $j$-$k$ for node
$i$.  The examples of $f_1(i, j\mbox{-}k)$ is shown in
Fig.~\ref{fig:edgeImportance}.  We assume $f_1(i, j\mbox{-}k) = f_1(i,
k\mbox{-}j)$ (symmetricity) and divide two cases of $f_1(i,
j\mbox{-}k)$ as follows.
\begin{itemize}
\item If two nodes $j$ and $k$ have the same distance from $i$, i.e.,
  $d(i, j) = d(i, k)$, then $f_1(i, j\mbox{-}k) = 0$.  (left of
  Fig.~\ref{fig:edgeImportance})
\item If two nodes $j$ and $k$ have the different distance from $i$,
  i.e., $d(i, j) + 1 = d(i, k)$, then $0 < f_1(i, j\mbox{-}k) \le 1$.
  We define node set $J$, each of which has an edge to $k$ and its
  distance from $i$ is equal to $d(i, j)$.
  \[
  J= \{ j' | j' \in V \mbox{ and } d(i, j') = d(i, j) \mbox{ and }
  d(j', k) = 1 \}
  \]
  Using $J$, $f_1(i, j\mbox{-}k)$ is defined as follows.
  \[
  f_1(i, j\mbox{-}k) = \frac{1}{|J|}
  \]
  The center of Fig.~\ref{fig:edgeImportance} shows a subcase of $|J|
  = 2$, and the right of it shows the other subcase of $|J| = 1$.
  Note that, this case includes the case of $i = j$.  If $i = j$, then
  $|J| = 1$ by the definition.
\end{itemize}

\subsection{Order of Edges Pairs for Local Search}

Two lower-importance edges are the candidate of swapping for 2-opt
local search.  All edges are sorted by edge importance and denoted by
$e_0, e_1, \ldots, e_{|E| - 1}$.  The edge $e_0$ has the smallest
importance.

The 2-opt local search algorithm is shown in Algorithm~\ref{alg:2opt}
The loop of lines~\ref{alg2opt:repoeat1} to \ref{alg2opt:repoeat1end}
is the main loop of local search.  Line~\ref{alg2opt:selectEdgePair}
is the important point using edge importance.  We tries several orders
to select a pair, which are described in the next paragraph.  For pair
$e_i (= a$-$b)$ and $e_j (= c$-$d)$ selected in
line~\ref{alg2opt:selectEdgePair}, there are two
combinations\footnote{If graph $G$ already has one of edges $a$-$c$ or
  $b$-$d$ ($a$-$d$ or $b$-$c$), we skip the swap since the degree of
  two nodes are decreased by one.} of swapping, 1) $a$-$c$ and $b$-$d$
($G'$ of line~\ref{alg2opt:G'}) and 2) $a$-$d$ and $b$-$c$ ($G''$ of
line~\ref{alg2opt:G''}) .  Diameter and ASPL of both $G'$ and $G''$
are calculated.  The loop of lines~\ref{alg2opt:while1} to
\ref{alg2opt:while1end} implies that edge importance is reused for
swapped graphs.  In our experience, after fifty swaps, ordering still
valuable to find smaller ASPL graph.
\begin{algorithm}
  \caption{2-opt local search}
  \label{alg:2opt}
  \begin{algorithmic}[1]
    \Procedure{Multiple2opt}{$G$}
      \State Calculate edge importance of all edges in $G$
      \State Sort edges by edge importance
      \While{timeout}\label{alg2opt:while1}
        \Repeat\label{alg2opt:repoeat1}
          \State Select an edge pair $(e_i, e_j)$ in a particular order\label{alg2opt:selectEdgePair}
          \State Generate swapped graph $G'$\label{alg2opt:G'}
          \State Calculate diameter and ASPL of $G'$
          \State Generate another swapped graph $G''$\label{alg2opt:G''}
          \State Calculate diameter and ASPL of $G''$
        \Until{$G'$ or $G''$ has smaller diameter or ASPL than $G$}\label{alg2opt:repoeat1end}
        \State Copy $G'$ or $G''$ to $G$
      \EndWhile\label{alg2opt:while1end}
    \EndProcedure
  \end{algorithmic}
\end{algorithm}

We heuristically employ two searching orders of
line~\ref{alg2opt:selectEdgePair} in Algorithm~\ref{alg:2opt}.  Both
orders satisfy $(e_i, e_j) < (e_i, e_k) < (e_j, e_k)$ for $i < j < k$.
We think the order of the smallest first is better than it of
triangle, empirically.
\begin{itemize}
\item the smallest first: $(e_i, e_l) < (e_j, e_k)$ for $i < j < k < l$.
  
  $(e_0, e_1), (e_0, e_1), (e_0, e_2), \ldots, (e_0, e_{|E| - 1}),$ \\
  $(e_1, e_2), (e_1, e_3), (e_1, e_4), \ldots, (e_1, e_{|E| - 1}),$ \\
  $(e_2, e_3), (e_2, e_4), (e_2, e_5), \ldots, (e_2, e_{|E| - 1}),$ \\
  $\ldots, $\\
  $(e_i, e_{i+1}), (e_i, e_{i+2}), (e_i, e_{i+3}), \ldots, (e_i, e_{|E| - 1}),$\\
  $\ldots$
\item triangle: $(e_i, e_l) > (e_j, e_k)$ for $i < j < k < l$.

  $(e_0, e_1),$ \\
  $(e_0, e_2), (e_1, e_2),$ \\
  $(e_0, e_3), (e_1, e_3), (e_2, e_3),$\\
  $\ldots, $ \\
  $(e_0, e_i), (e_1, e_i), (e_2, e_i), \ldots, (e_{i - 1}, e_i),$ \\
  $\ldots$
\end{itemize}

Since ASPL calculation is time-consuming task, we additionally design
ASPL recalculation method for 2-opt local search.  The method stores
the distance matrix of graph $G$ and update the matrix for swapped
graph $G'$.  We can elimiate re-calculation of the distance between
nodes which the swap does not affect.

\subsection{Graph Instances}

We run local search program during the competition and after
competition.  We show the smallest graph that we found in
Table~\ref{tbl:2optResults}.  These graphs probably are the best-known
graphs for these four combinations of order and degree.  For graphs of
order $n = 4096$ and $n = 10000$, Algorithm~\ref{alg:2opt} is directly
applied.

\begin{table}
  \caption{Graphs by local searching after 2015 competition}
  \label{tbl:2optResults}
  \centering
  \begin{tabular}{rr|rr|r}
    order $n$ & degree $d$ & diameter $k$ & ASPL $l$ & $l$ of Table~\ref{tbl:heuristicResults} \\
    \hline
    256 & 16 & 3 & 2.09069 & 2.12757 \\ 
    4096 & 60 & 3 & 2.295216 & 2.295275 \\ 
    4096 & 64 & 3 & 2.242170 & 2.242228 \\ 
    10000 & 60 & 3 & 2.648977 & 2.648980 \\ 
    \\
    \multicolumn{5}{r}{\begin{minipage}{.7\linewidth}

        Note: all graphs may be the best-known graphs.  The winner's
        graph of the competition has larger ASPL than these.
    \end{minipage}}
  \end{tabular}
\end{table}

\begin{figure}[tb]
  \begin{center}
    \includegraphics[width=\linewidth]{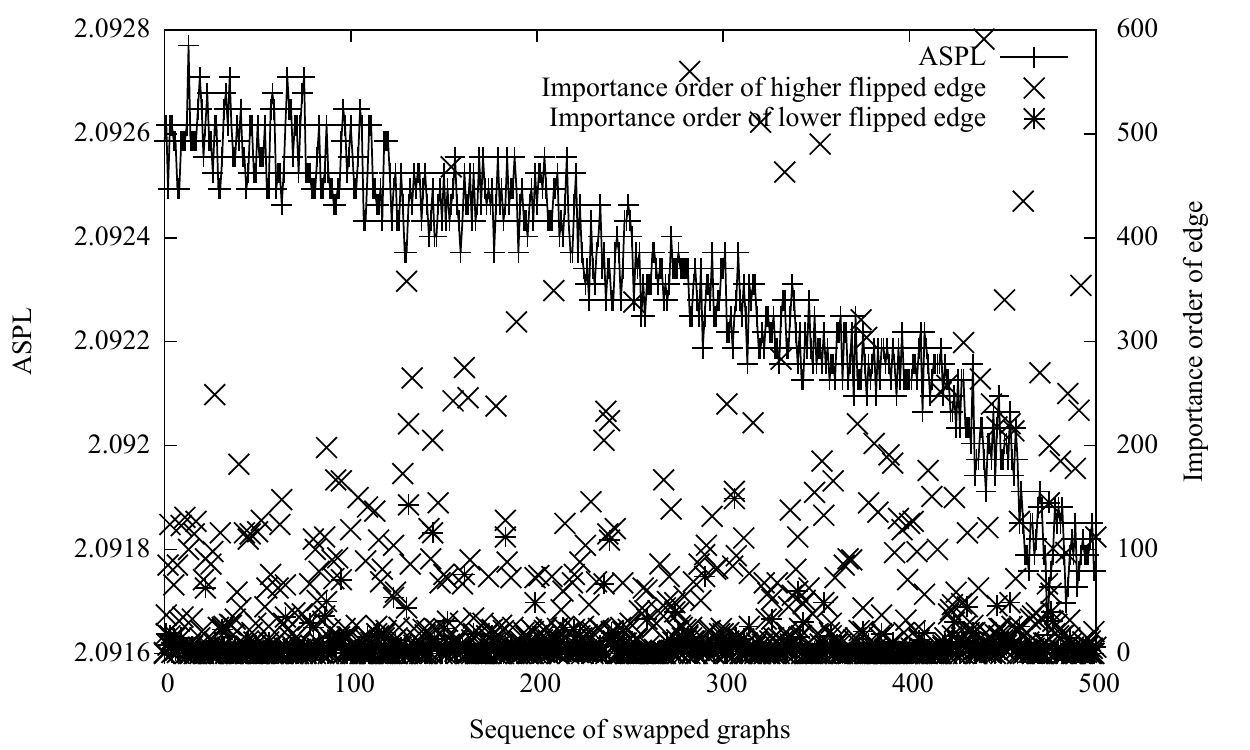} 

    \footnotesize
    (a) first half (ASPL: $2.09163 < l < 2.09277$)
    
    \includegraphics[width=\linewidth]{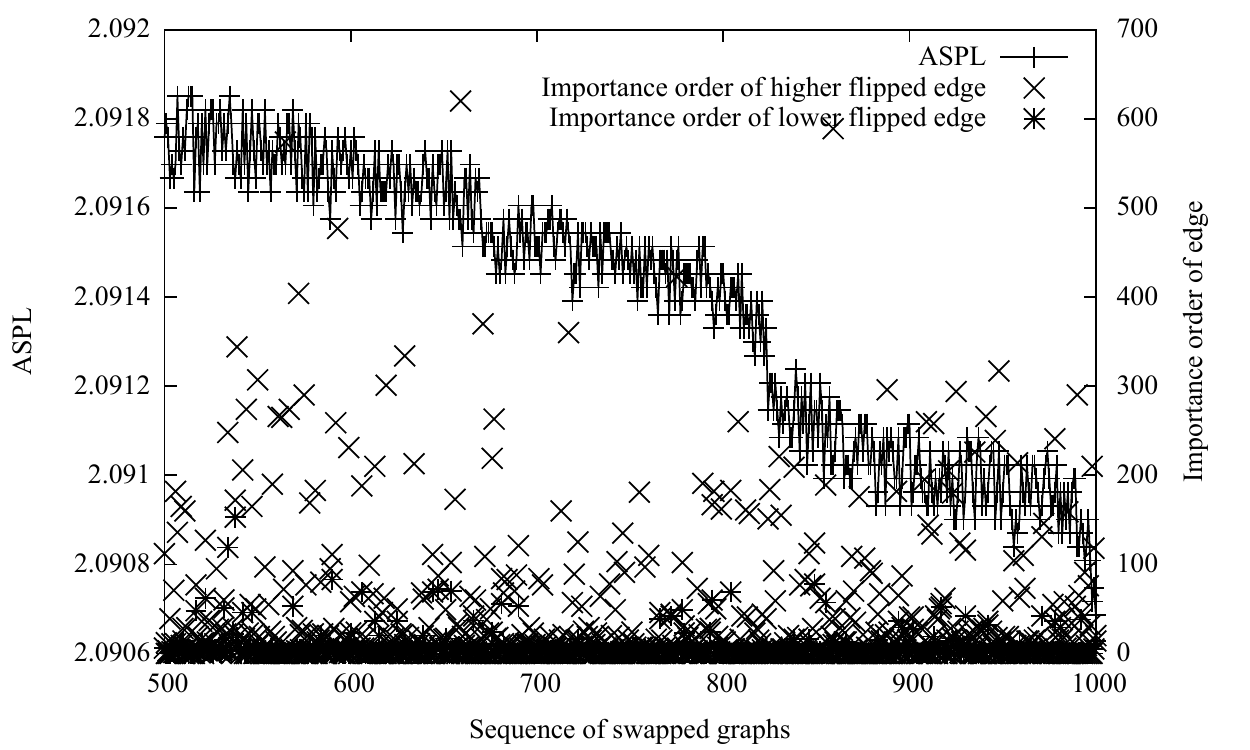} 

    (b) last half (ASPL: $2.09068 < l < 2.09186$)
  \end{center}
  \caption{Local search history of 1000 graphs of order $n = 256$ and degree $d = 16$}
  \label{fig:localsearch-n256-d16}
\end{figure}

For the graph of order $n = 256$ and degree $d = 16$, our graphs fall
into local optimal many times.  To find graphs of smaller ASPL, we
accept worse post-graph $G'$ than pre-swap graph $G$ in 2-opt local
search.  Fig.~\ref{fig:localsearch-n256-d16} shows the search history
of the last 1000 graphs before reaching the best-known graph of $l =
2.09069$.  Many branches from each graph is omitted.  In this figure,
we show ASPL of each graph and order of edges that we swapped.  The
order of swapped edges is distributed from 0 to 620, i.e., swapped
edges are two of $e_0, e_1, e_2, \ldots, e_{620}$ in each graph.  To
reach the best-known graph, we need to run local search at least in
the range of edge pair $(e_i, e_j)$ for $0 \le i \le 153$ and $1 \le j
\le 620)$.  This range contains only 4 \% of all edge pairs.  So, the
edge importance seems to be valuable function to prioritize edges for
swapping.
We briefly explain the distribution of order of swapped edges.  Since
two edges are selected for each swap in
Fig.~\ref{fig:localsearch-n256-d16}, the total number of selected
edges are 2000 for 1000 swaps.  The half of these edges have the order
smaller than 8.  The order smaller than 108 contains 90\% of these
edges.


\section{Conclusion}

In this paper, we explained the heuristic algorithm that creates a
graph which has small average shortest path length (ASPL) for diameter
3 graphs.  The algorithm intends to increase the number of pentagons
(5-node cycles).  Through the observation of small order graphs which
has diameter 3, we focused on the number of pentagons.  The heuristic
algorithm can create two best-known graphs at the graph golf 2015
competition, and a best-known graph after the competition.  These
three graphs have order $n = 4096$ and degree $d=60$, $n = 4096$ and
$d=64$, and $n = 10000$ and $d=60$.

We also explained the technique of 2-opt local search to reduce ASPL
of a graph.  The technique is based on the evaluation function called
edge importance (or edge impact).  Edges which have smaller importance
compared to other edges are the good candidates for swap of 2-opt.  We
applied this technique to the graph of $n = 256$ and $d = 16$, and
find a best-known graph after competition.

As future work, we will try to find more elegant heuristic to create a
small ASPL graph, for not only diameter 3 graphs but also larger
diameter graphs.  Data structures for fast ASPL computation also
should be explored.


\section*{Acknowledgment}
This work was partially supported by JSPS KAKENHI Grant Number
26330107.



%

\end{document}